\documentclass[structabstract]{aa} 
\usepackage{graphicx}
\usepackage{txfonts}
\begin{document}

   \title{High-resolution infrared spectroscopy as a tool to detect
          false positives of transit search programs
\thanks{partly based on observations obtained at the European Southern
    Observatory at Paranal, Chile in program 081.C-0033(A).}}

   \author{Eike W. Guenther
          \inst{1,2}
          Lev Tal-Or 
          \inst3}
   \institute{Th\"uringer Landessternwarte Tautenburg,
              Sternwarte 5, D-07778 Tautenburg, Germany\\
              \email{guenther@tls-tautenburg.de}
         \and
                Instituto de Astrof\'\i sica de Canarias,
                C/V\'\i a L\'actea, s/n,
                E38205 -- La Laguna (Tenerife), Spain
         \and
                School of Physics and Astronomy, 
                Raymond and Beverly Sackler Faculty of Exact Sciences, 
                Tel Aviv University, Tel Aviv, Israel
            }
             
   \date{Received November 15, 2009; accepted November 16, 2009}

   \abstract 
   {Transit search programs such as CoRoT and Kepler now have the
     capability of detecting planets as small as the Earth. The
     detection of these planets however requires the removal of all
     false positives. Although many false positives can be identified by
     a detailed analysis of the light-curves (LCs), the detections of 
     others require additional observations. An important source of 
     false positives are faint eclipsing binaries within the point 
     spread function (PSF) of the target
     star. For example, triple stars are an important source
     of false positives. Unfortunately, most of the methods previously
     applied have difficulties in detecting these objects.}
  { We develop a new method that allows us to detect faint eclipsing
    binaries with a separation smaller than one arcsec from target
    stars. We thereby focus on binaries that mimic the transits of
    terrestrial planets. These binaries can be either at the same
    distance as the target star (triple stars), or at either larger,
    or smaller distances.}
   { A close inspection of the problem indicates that these systems
     contain either late-type stars, or stars of high
     extinction. Thus, in both cases the binaries are brighter in the
     infrared than in the optical regime. We show how high resolution
     infrared (hereafter IR) spectroscopy can be used to remove these
     false positives.}
   { For the triple star case, we find that the brightness difference
     between a primary and an eclipsing secondary is about 9-10 mag in
     the visual but only about 4.5-5.9 magnitudes in the K-band.  In
     the next step, we demonstrate how the triple star hypothesis can
     be excluded by taking a high-resolution IR spectrum. Simulations
     of these systems show that the companions can be detected with a
     false-alarm probability of $\sim 2\%$, if the spectrum has a
     signal-to-noise ratio (S/N) $\geq$ 100. We subsequently show that
     high-resolution IR spectra also allows to detect most of the
     false positives caused by foreground or background binaries.}
   { If high resolution IR spectroscopy is combined with photometric
     methods, virtually all false positives can be detected without
     radial velocity (RV) measurements. It is thus possible to confirm transiting
     terrestrial planets with a modest investment of observing time.}
   \maketitle
%

\section{Introduction}

Space-based programs such as CoRoT and Kepler searching for transiting
planets now have the capability of detecting planets as small as
one or a few Earth-radii (Leger et al. \cite{leger09}; Queloz et al.
\cite{queloz09}). An outstanding problem however, is the detection of
false positives, i.e. objects that have light-curves resembling those
of a transiting planet but which are something else. Excluding these
false positives is an essential part of any transit search program,
particularly those focusing on very small planets.

In a detailed study, Brown (\cite{brown03}) investigated the different
possibilities for false-alarms, and concluded that diluted binaries
are the dominant sources of false positives for space-based transit
search programs. These can be either a foreground or background
eclipsing binaries too close to the target to be resolved.  Triple
systems of small separation can be considered as a special case of a
diluted binary. For the Kepler mission Gautier et
al. (\cite{gautier07}) and Caldwell et al. (\cite{caldwell06})
demonstrated that diluted binaries can be detected if they are
separated from the target star by more than a quarter of a pixel,
corresponding to about one arcsec on the sky.  

For the CoRoT mission, Almenara et al. (\cite{almenara09}) showed that
83\% of the transit candidates are identified as false positives by
carefully analysing the light-curves.  In the case of CoRoT, an
important way to discriminate eclipsing triple systems from transiting
planets is to use the colour difference in and out of transit (see
also O'Donovan et al. \cite{odonovanl06}; Tingley
\cite{tingley04}). This method was also used for \object{CoRoT-7b} but
the S/N ratio in the blue and green channel alone was found to be
insufficient to exclude a late type M-star (Leger et al.
\cite{leger09}). This is because it is not only necessary to detect
the transit in both colours but to measure the depth of the transit
with sufficient accuracy.  Furthermore the study by Almenara et
al. (\cite{almenara09}) show that another 5\% of the transit
candidates found by CoRoT are planets, and the remaining 12\% are
false positives that require additional observations.

The efficiency in removing false positives is so high because CoRoT
performs three colour photometry. Follow-up observations thus
concentrate on only 17\% of the candidates, 30\% of which are
planets. The first step of the follow-up program is to identify
possibly contaminating stars in the Exo-Dat database (Deleuil et
al. \cite{deleuil09}), and observe them in and out of transit (Deeg et
a. \cite{deeg09}). Although these methods allow us to remove many
false positives, some remain.

The most important remaining case is a faint eclipsing binary with a
separation of smaller than an arcsec from the target.  These eclipsing
binaries may be either in the background, the foreground, or at the
same distance as the target star, as in a triple system.

In this article, we discuss a technique for identifying false
positives by detecting the infrared CO lines of eclipsing binary M
stars in the combined light of blended systems.  This important,
because methods based on RV-measurements require much observing time
to detect planets. For example, 106 RV measurements were necessary to
determine the mass of CoRoT-7b, a planet of $4.8\pm0.8$ $M_{Earth}$
and an orbital period of 0.8535 days (Queloz et al \cite{queloz09}).
In the case of a very low-mass planet in a long period orbit, it might
be entirely impossible to confirm it by means of RV measurements.  For
example, the semi-amplitude of an earth-like planet orbiting at 1 AU
is only 0.09 $m\,s^{-1}$.  It is thus important to develop methods
that allow to us verify transiting extrasolar planets without RV
measurements.  The method assumes that differences in RV -- either
from orbital motion in triple systems or simple line-of-sight
differences -- separate the spectral lines in wavelength space.

\section{Excluding triple systems}

In this section we describe how triple systems can be excluded as
source of false positives by means of high-resolution infrared
spectroscopy.

\subsection{The brightness difference between the
primary and secondary is smaller in the IR}


As outlined above, triple systems may be a source of false positives
but how relevant are they?  Since the CoRoT and Kepler mission focus
on main-sequence F, G, and K stars ($M_{*}=0.5$ to 1.7 $M{_\odot}$),
only these stars are discussed in the following. According to
Tokovinin (\cite{tokovinin08}) at least 8\% of the solar-type stars
are in systems containing three or more stars.  Dynamically stable
triple systems have to be hierarchical with $P_{L}/P_{S}>5$ (Eggleton
\cite{eggleton06}).  This means that even for an eclipsing system with
an orbital period of one year, the separation between the primary and
the eclipsing system can be as small as 3.0 AU.  If we consider the
canonical 12th magnitude G-star, this means that for the given
semi-major axis, the ratio of the ratio of semi-major axis to distance
can produce a maximum separation as small as 0.01 arcsec.

Using the results obtained by Tokovinin (\cite{tokovinin08}), we find
that the probability of a triple system containing two M-stars with an
orbital period of one year is between 0.1 and 0.4\%.  The probability of a
triple system containing two eclipsing M-stars with an orbital period
of up to one year is $\geq 10^{-4}$.  Since the CoRoT and Kepler 
both survey $10^{5}$ stars, triple systems are an important source of
false positives. It is thus unsurprising that CoRoT has already
found at least one such system.

The positives aspect of a triple star is that all three components are
at about the same distance from the Earth. Thus, given the depth of
the transit, the spectral type of the primary, and the period of the
transit, all relevant parameters of a hypothetical triple star system
can be fixed. For example, an eclipsing binary that mimics a
transiting planet such as the Earth would have to consist of two
M-stars.  The most interesting case is a transit that mimics a planet
such as the Earth.

What would be the properties of a triple star causing such a LC? The
triple star will consist of a F, G, K-star primary and eclipsing M-star
binary.  The two relevant cases for the M-star binary are:

\begin{itemize}
\item An eclipsing system consisting of two identical stars,
  such as \object{YY Gem}. In this case, the primary and secondary
  eclipse would have the same depth.
\item A eclipsing system where the secondary eclipse is too
  shallow to be detected in the analysis of the LC. These are systems
  in which one component is much fainter than the other.
\end{itemize}

Unequal mass (M-star) binaries for which the secondary eclipse is
detectable do not have to be considered here, as these false positives
would already be removed in the analysis of the LC. There are two
other less relevant cases. One is that $i, \Omega$, and $e$ are such
that the primary eclipse is observed but not the secondary. We do not
have to discuss this case separately, because the method applied to
remove this case is identical to when one component is much fainter
than the other. The method for excluding these two cases are
identical.

Another possibility is a binary with an eccentric orbit in which the
impact parameter is such that two transits of equal duration and depth
are produced by stars of different radii. This case is very
unlikely. Since the method for excluding this case is the same as for
the case of identical stars, we do not have to discuss this case
separately either. We thus have to exclude only the two relevant cases
given above.

Transits of binaries can be either central or grazing. For a given
transit depth, the eclipsing stars are brighter if the transit is
grazing rather than central. Since binaries with central eclipses are
more difficult to detect, we focus on central eclipses. In additional,
all values are given for a transiting planet as the Earth but can be
scaled to larger or smaller planets.

Since the depth of the transit is simply the squared of the ratio of
the radii, the depth depends not only on the size of the planet but
also the size of the star.  The transit is shallower if the star has a
larger diameter, and deeper if the star has a smaller diameter. The
depth of the transit of a terrestrial planet is $5.8\times10^{-5}$ for
an F5V-star, $9.6\times10^{-5}$ for a G5V-star, and $1.5\times10^{-4}$
for a K5V star.

\begin{figure}[h]
\includegraphics[width=0.38\textwidth,angle=-90]{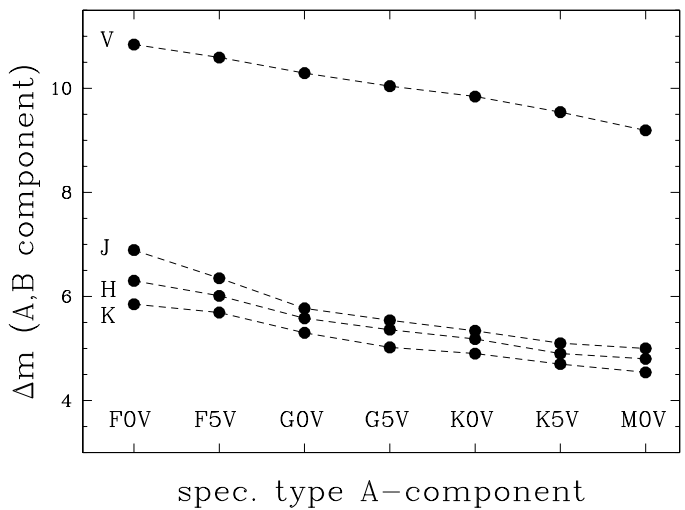}
\caption{Maximum difference in brightness between the A and the B
  component in mag ($\Delta m_{V},\Delta m_{J},\Delta m_{H},\Delta$
  ).}
\label{diff}
\end{figure}

What would be the properties of an eclipsing binary that mimics the
transit of a terrestrial planet? In the case of an F5V star, the
brightness-ratio of the primary to the secondary (the brighter of the
two eclipsing stars) would be $I_{sec}/I_{prim}\leq
5.8\times10^{-5}$. This corresponds to a brightness-difference
$m_{sec}-m_{prim}\leq 10.6$ mag in the visual. Figure\,\ref{diff}
shows the maximum difference in mag between primary and the secondary
in the V,J,H,K-band ($\Delta m_{V},\Delta m_{J},\Delta m_{H},\Delta
m_{K}$) for main-sequence stars of different spectral types. The
brightness difference between the hypothetical companion and the
primary depends on the spectral type of the primary star, because the
depth of the transit depends on the size of the host star.  The
brightness difference is thus smaller for a main sequence K-star than
for an F-star (again for a transit of a planet of the same size as the
Earth).

At first glance, we may consider it straightforward to acquire a
high-resolution spectrum in the optical regime with a sufficient S/N
to detect the companion. The spectral lines of the primary and the
eclipsing binary would be well separated from both each other and the
primary. For example, the semi-amplitude of an eclipsing binary
consisting of two 0.09 $M\odot$-stars with an orbital period of one
year is 18 $km\,s^{-1}$. However, the problem is that the brightness
difference in the optical is about 10 mag. To achieve a detection we
would thus require a S/N of 10 000. As we now show, the situation is
more promising in the IR.

To estimate the brightness of the hypothetical eclipsing binary, we
can assume that all three stars are at about the same distance from
Earth. For a given brightness difference, we can thus calculate the
spectral-type of the hypothetical companion. We find that the
hypothetical companion would have a spectral-type in the range from
M4V to M7V, if we were to assume a central transit. Using these
hypothetical-companion spectral-types, we can derive the brightness
differences for any other spectral region using the information given
in Henry et al. (\cite{henry06}). For a grazing transit, the
spectral-types would be earlier and the brightness difference
correspondingly smaller.

As expected, the brightness difference between a main-sequence F, G, 
or K-star and an M-star is much smaller in the IR than in the optical
regime.  The brightness difference in the K-band is only $5.7\pm0.2$
mag for a main sequence F-star, $5.0\pm0.2$ mag for a G-star, and
$4.8\pm0.2$ mag for a K-star, respectively (Fig.\,\ref{diff}).

\begin{figure}[h]
\includegraphics[width=0.38\textwidth,angle=-90]{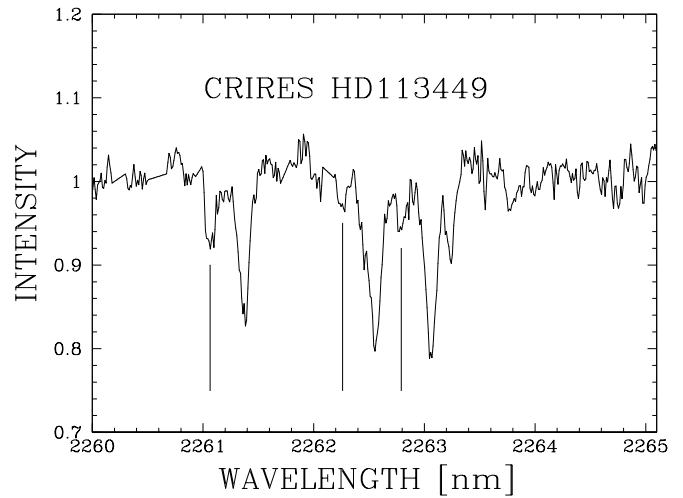}

\caption{CRIRES spectrum of the binary HD113449 consisting of a
  G5V primary and an M3V secondary. Spectral lines of the secondary are
  marked.}
\label{hd113449}
\end{figure}

Since it has already been demonstrated that faint late-type
secondaries of binaries can be efficiently detected by using
high-resolution IR spectroscopy, we can also use this method
to detect false positives (Mazeh et al. \cite{mazeh02}).  A practical
demonstration that an M-star companion of G-star can be detected is
shown in Fig.\,\ref{hd113449}. The figure shows a CRIRES spectrum
\object{HD113449} consisting of a G5V primary and an M3V
secondary. Spectral lines of both components are clearly visible.

\subsection{Using the CO-lines in the infrared}

In Sect. 2.1, we showed that the hypothetical companions are
in the limiting cases M4V to M7V-stars, and pointed out that
high-resolution IR spectroscopy can be used to detect these
false positives. We now discuss the instrumental
requirements, illustrating how it works in practice.

The most effective way to exclude an M-star companion is to observe a region in
the spectrum, in which the spectral lines of an M-star are stronger
than the lines from an F, G, or K star. For example we
consider that a spectral-line is absent in the primary but has a
depth of 100\% in the secondary, and that the secondary would be 5 mag
fainter. In this case, a spectrum with a S/N of 100 would be required
to detect the secondary. If the line were only 10\% deep, we
would need a spectrum with a S/N of 1000. However, if the secondary
exhibited 100 of these lines and we used the cross-correlation method to
average them, we could still detect the secondary using a spectrum  
with a S/N of 100.

\begin{figure}[h]
\includegraphics[width=0.38\textwidth,angle=-90]{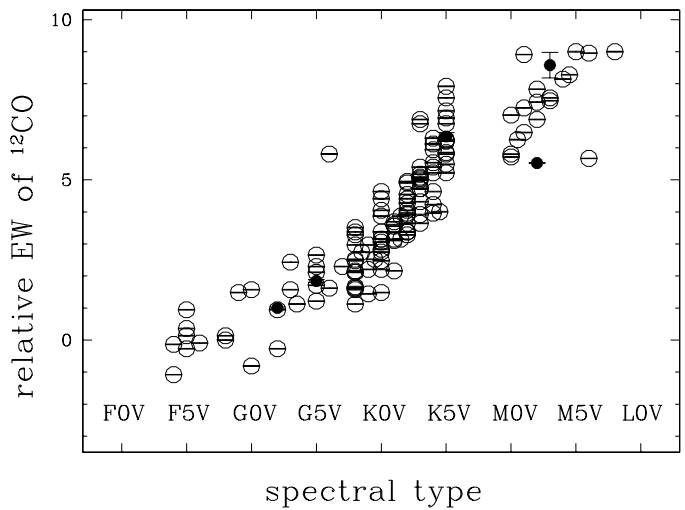}
\caption{Equivalent width of the $^{12}CO$ lines in the K-band
relative to the strength of these lines in a G2V star.}
\label{CO}
\end{figure}

Because many high-resolution IR spectrographs have a rather limited
wavelength coverage, we now discuss which part of the spectrum
would be most suitable. For three reasons, the CO overtone lines 
(2.38 to 2.45 $\mu m$) seem to be the most suitable set of lines in 
the J, H, and K-bands:

\begin{itemize}
\item CO has a large number of deep lines ($\sim$ 50 lines). 
\item The overtone lines are in the K-band, where the brightness-difference 
      between an M-star and a G-star is smaller than in the J or H-band.
\item The lines are much stronger for an M-star than for an F, G, or
      K-type star (Fig.\,\ref{CO}).
\end{itemize}

Goorvitch (\cite{goorvitch94}) provides a list of 50 $^{12}C^{16}O$ lines
in the K-band. The atlas published by Wallace, Hinkle, \& Livingston
(\cite{wallace01}) shows that these lines are typically 50\% deep in a
sun-spot (corresponding to an M3V star).

\begin{figure}[h]
\includegraphics[width=0.38\textwidth,angle=-90]{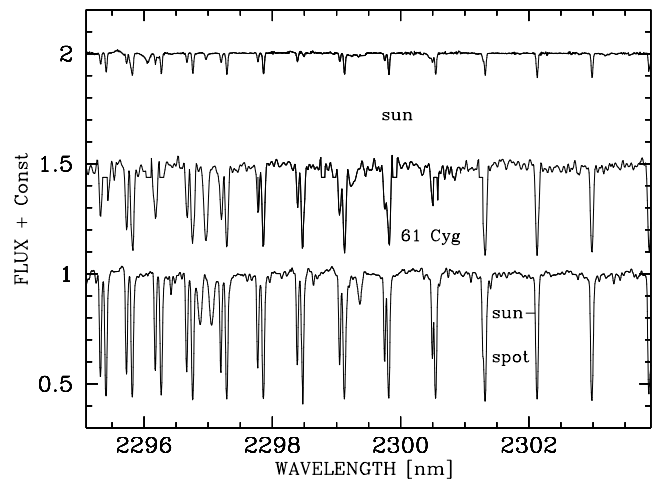}
\caption{ Spectra of the Sun (G2V), 61 Cyg B (K7V), and a sunspot (M3V),
  demonstrating the increase in the strength of the CO-lines towards
  later spectral types.  }
\label{COlines}
\end{figure}

The third reason given above requires some additional discussion.  The
optimal way of quantifying the strength of the lines would be to use
high-resolution spectra of stars in the spectral-type range between
early F and late M.  High-resolution spectra of the Sun, (G2V),
\object{61 Cyg B} (K7V), and a sunspot (M3V) are shown in
Fig.\,\ref{COlines} (Livingston \& Wallace \cite{livingston91};
Wallace \& Hinkle, \cite{wallace96b}; Wallace, Hinkle, \& Livingston
\cite{wallace01}).  Unfortunately, until now there has been an
insufficient number of high-resolution spectra to quantify the line
depth for all spectra types but sufficient low- and medium- resolution
spectra. The strength of the lines is often expressed in terms of the
$^{12}CO(2,0)$-bandhead index. For example, M{\'a}rmol-Queralt{\'o} et
al. (\cite{marmol08}) developed a new CO-index that is less dependent
on the resolution of the spectrograph. This index can thus be used to
quantify the strength of the lines using low and medium resolution
spectra. To calibrate this CO-index, all we have to do is to measure
the line depth in a few very high resolution spectra.  For this
purpose, we use FTS spectra published by Wallace et
al. (\cite{wallace96a}) and Wallace et al.  (\cite{wallace96b}), and
scale the CO-index values accordingly.  Figure\,\ref{CO} shows the
strength of the $^{12}CO$-lines for different spectral types relative
to the sun.

\begin{figure}[h]
 \includegraphics[width=0.38\textwidth,angle=-90]{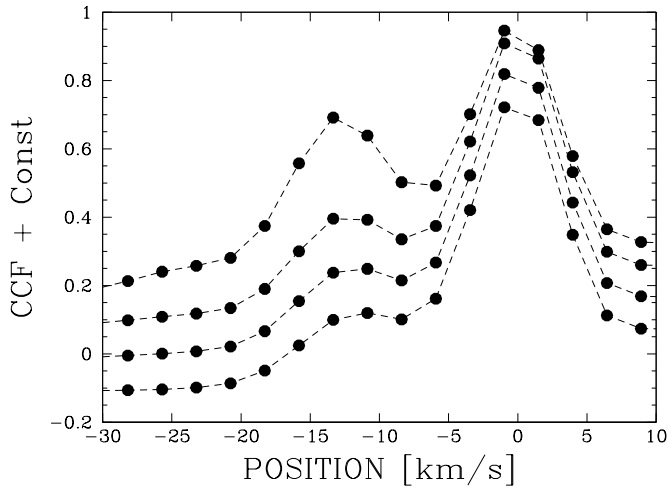}
 \caption{Cross-correlation function for a solar-like star
 and an M-star companion which is 3, 4, 4.5 and 5,0 mag
 fainter.}
 \label{CCF}
 \end{figure}

In good agreement with the older data (Kleinmann \& Hall
\cite{kleinmann86}), the new measurements show that the strengths of
the CO-lines increase for stars with spectral types up to M6V. For
later spectral types, the strengths decrease again (Cushing et
al. \cite{cushing05}), but this is not relevant to the case of the
triple system. We discuss this case in the case of a foreground
binary. Figure\,\ref{CCF} shows the appearance of
the cross-correlation function for companions that are 3, 4, 4.5, and 
5 mag fainter, and shifted by $13\,km\,s^{-1}$.

We now derive the equation for the S/N required to exclude a
binary mimicking a planet of a given radius.  We assume that there
are at least 50 CO-lines within the spectral range observed, and that the 
resolution is high enough to resolve them
($\lambda/\Delta \lambda \geq 30000$). The S/N that is required is

$$ S/N \geq 100 \cdot 2.512^{(5-\Delta m_{K})} \cdot {{8}\over{RelCO}}
\cdot {{R_{Earth}}\over{R_{planet}}},$$

\noindent
where $\Delta m_{K}$ is the brightness difference between the primary and
secondary, $\Delta m_{K}$ is 5.9-4.5 mag for F0V to M0V stars
(Fig.\,\ref{diff}),  $R_{planet}$ is the radius of the planet that is
to be confirmed, and $RelCO$ is the relative strength of the CO-line
shown in Fig.\,\ref{CO} ($RelCO\sim8$ for M-stars). Using this
equation, we conclude that a $S/N\sim 100$ is sufficient for identifying
this kind of false positive.

\begin{figure}[h]
\includegraphics[width=0.38\textwidth,angle=-90]{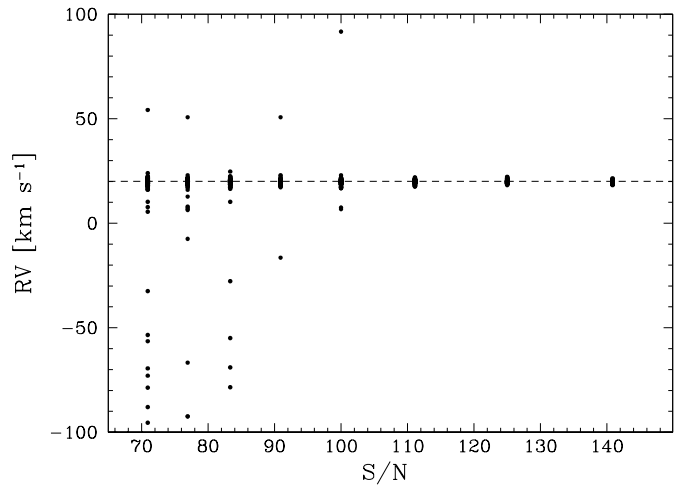}
\caption{RV-measurements obtained with TODCOR for a simulated binary
  consisting of a G2V primary and an M-star secondary that is 5 mag
  fainter in the K-band for different S/N-ratios.}
\label{TODCOR}
\end{figure}

With TODCOR we now demonstrate that the detection is possible.  TODCOR
is a 2-dimensional cross-correlation method that has already been
indeed applied to detect very faint companions in IR spectra (Mazeh et
al. \cite{mazeh02}). We performed simulations in which we used the FTS
spectra (Livingston \& Wallace \cite{livingston91}; Wallace \& Hinkle
\cite{wallace96b}) and resampled them to a resolution of
$\lambda/\Delta \lambda = 60\,000$ to mimic CRIRES observations with a
0.4 arcsec slit. Our simulated binary consisted of a G2V primary and
an M-star companion, which was shifted by $20\,km\,s^{-1}$, was 5
magnitudes fainter in the K-band.  We finally added random noise to
the spectra and carried out 1300 simulations with different
S/N-ratios. For a S/N-ratio of 100, we performed 200 simulations.
Given that there are three outliers in all our simulations, the
false-alarm probability is $\sim 2 \%$ for a S/N of 100. These
simulations thus confirm our expectation.  Figure\,\ref{TODCOR} shows
the RV measurements obtained with TODCOR for different values of S/N.
For higher S/N the false-alarm probability is lower, but increases
rapidly towards lower S/N. For example, at a noise-level of 85, the
false-alarm probability is already $5\%$.  To exclude false-alarms of
terrestrial planets, we thus recommend using spectra with a $S/N \geq
100$.

In principle, it could be argued that the fundamental bands are more
suitable than the overtone-bands, since they are about a factor six
stronger (Heras et al. \cite{heras02}).  However, these lines are in
the wavelength range between 4.30 and 4.70 $\mu m$, thus almost
unobservable from the ground.

\section{Excluding a binary in the background}

Another possibility to be excluded is that there is an unrelated eclipsing
binary in the background. Although most of the background binaries can
be removed by analysing the light-curves not all them can be removed
in this way.  The usual methods have problems in removing binaries
that are faint and close to the primary. In the following,
we show that high-resolution IR spectroscopy also helps to remove these
false positives.  As before, the brightness difference can be as
large as 9.2-10.8 mag in the optical regime (Fig.\,\ref{diff}).  Would
the brightness difference be smaller in the IR?  Calculating the
brightness of the background star in the IR is not easy, because we do
not know its spectral type. We thus have to consider two cases:

\begin{itemize}
\item At least one of the stars is a giant star. 
\item The eclipsing background binary consists of two main-sequence stars.
\end{itemize}

\noindent
Since typical targets are between 10th and 14th magnitude, we have to
exclude there being background binaries as faint as 19th or even 25th mag in the
visual.

Giant stars in the field would certainly be much brighter in the IR
than in the optical regime. As shown in Deleuil et
al. (\cite{deleuil06}), about 70\% of the giant stars in the CoRoT
fields are K-stars, the remainder being G-type giants. Without
extinction, the V-K colours are in the range between 2.3 and 3.6 mag for
K-giants, and 1.8 and 2.2 mag for G-giants, respectively. If we take
into account that the surveys are carried out close to the plane of
the Milky Way, the extinction will make these objects even redder.
Brown (\cite{brown03}) finds that the J-K colours of background giant
stars are between 0.5 and 1.5 mag. Background giants are thus at
least two mag brighter in the K-band than in the optical regime.
Thus, acquiring IR spectra can greatly increase our likelihood of
detecting background giants. This is especially true for
K-giants, since K-stars have strong CO-line overtone lines.

\begin{figure}[h]
\includegraphics[width=0.38\textwidth,angle=-90]{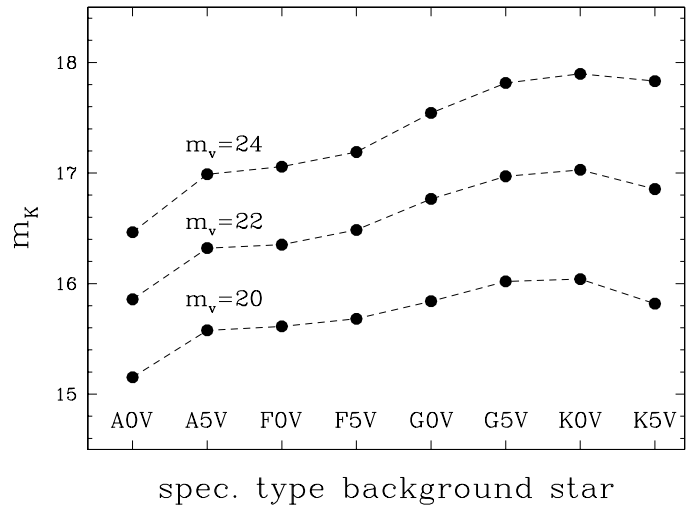}
\caption{Brightness of an unrelated main-sequence background star
  with $m_v=20,22,$ and 24 in the K-band. Because the extinction is
  smaller in the IR, the objects are much brighter in this
  wavelength regime.}
\label{back}
\end{figure}

Dust extinction also changes the colours of main-sequence stars.
Figure\,\ref{back} shows the brightness of background stars in the
K-band. To estimate the interstellar extinction, we used a model published by
Beatty \& Gaudi (\cite{beatty08}) and assumed that the observation was
carried out in the Galactic plane. Because the extinction is much
smaller in the K-band than in the visual, background stars are much
brighter in the IR than in the optical regime. Our estimates are
confirmed by the results of Deleuil et al. (\cite{deleuil06}), who show
that faint stars in the CoRoT fields have red colours.

Examples of background binaries that cannot be excluded using IR
spectroscopy are sdBs and cataclysmic variables because these objects are
blue. While cataclysmic variables cannot have orbital periods of a
year, this is possible for sdBs and white dwarf binaries.  Maxted,
Marsh, \& North (\cite{maxted01}) give a space density for sdBs of
$10^{-6}\,pc^{-3}$. The space density of sdBs is thus very low, and
this case is unlikely. If we were to take into account that a
sdB can only have blue colours if the extinction is low, the binary
would have a distance of about 100 kpc to be of $m_V=24$. Since we observe
in the plane of the Milky Way, this case is not possible. Thus, neither
sdBs nor CVs can cause false positives.

In addition to spectroscopy, IR photometry can be used to exclude
unrelated background binaries. If we find that the transit is deeper
in the IR, we can infer that it is a binary.

\section{Excluding a binary in the foreground}

We finally discuss the possibility of an unrelated eclipsing
binary in the foreground. As for the background objects, the
brightness difference can be as great as 10.8 mag in the visual. A
binary in the foreground that is 10.8 mag fainter has a spectral type
later than M6V, unless it consists of white dwarfs, in which case it could be a
M, L, T type binary.  This object would again be much brighter in
the IR than in the optical.  To demonstrate this, we assume a
binary consisting of two objects such as \object{2M1507-16}, which
has a spectral type L5V, and brightnesses of
$m_v=22.9\pm0.5$, $m_k=11.4\pm 0.2$ (Dahn et al. \cite{dahn02}).  This
means that a binary consisting of two of these objects would be as bright as
the primary in the IR! It would certainly be very easy to exclude this
case. At M9.5, $V-K$ is 9.3 mag, making the detection of a
foreground binary in the IR easier than either a triple system or a
background binary.

We can not exclude using IR spectroscopy the presence of white dwarf
binary. Maxted, Marsh, \& North (\cite{maxted01}) calculate a space density
of white dwarf binaries of $5\times 10^{-3}\,pc^{-3}$. The probability of
finding an unrelated eclipsing white dwarf system with an orbital
period of about a year within one arcsec from the target star is
$<10^{8}$ and thus irrelevant.

\section{Conclusions}

Most of the methods that have been used to detect
false-positives in transit search programs have difficulties in
detecting faint eclipsing binaries within the PSF of the target star.
Examples of these binaries triple stars containing two eclipsing
M-stars. The problem is that the separations of the stars in these
systems can be too small to be spatially resolved.  Since triple stars
are common, they are an important source of false positives. Another
source being eclipsing background binaries. We 
have demonstrated that both types of false positives can be effectively
eliminated using high resolution IR spectroscopy. The reason 
for this is that eclipsing stars either are at large distances and thus suffer
from extinction, or are nearby late-type stars.  Thus, in both
cases, they are brighter in the IR than in the optical
regime. If high resolution IR spectroscopy is combined with
photometric methods, virtually all false positives can be detected
without RV measurements. It is thus possible to confirm transiting
planets with a modest investment of observing time.

\begin{acknowledgements}
      EG would like to thank Stephan Geier for his information
      about sdBs and CVs.
\end{acknowledgements}

\end{document}